\documentclass[numberedappendix]{emulateapj}
\usepackage{graphicx}
\usepackage{natbib}
\usepackage{amsmath}
\usepackage{color}
\usepackage{comment}
\usepackage{rotating}
\newcommand{\GALFIT}{\textsc{Galfit}~}
\newcommand{\sextractor}{\textsc{SExtractor}~}

\usepackage{eso-pic}

\AddToShipoutPictureBG*{%
  \AtPageUpperLeft{%
    \hspace{0.75\paperwidth}%
    \raisebox{-10.5\baselineskip}{%
      \makebox[0pt][l]{\textnormal{FERMILAB-PUB-16-450-AE}}
}}}%

\bibpunct{(}{)}{;}{a}{}{,}

\begin{document}
\title{M32 Analogs? \\ A Population of Massive Ultra Compact Dwarf and Compact Elliptical Galaxies in intermediate redshift CLASH Clusters} 

\author{Yuanyuan. Zhang$^{1, 2, \dagger}$ \& Eric F.\ Bell$^{3}$}
\affil{$^1$Fermi National Accelerator Laboratory, Kirk \& Pine Road, Batavia, IL 60510, USA}
\affil{$^2$Physics Department, University of Michigan, 450 Church Street, Ann Arbor, MI 48109, USA}
\affil{$^3$Department of Astronomy, University of Michigan, 1085 S.\ University Avenue, Ann Arbor, MI 48109, USA}
\email{\hspace{1em}$^\dagger$Email: ynzhang@fnal.gov}
\begin{abstract}

We report the discovery of relatively massive, M32-like ultra compact dwarf (UCD) and compact elliptical (CE) galaxy  candidates in $0.2<z<0.6$ massive galaxy clusters imaged by the Cluster Lensing And Supernova survey
with {\it Hubble} (CLASH) survey. Examining the nearly unresolved objects in the survey, we identify a sample of compact objects concentrated around the cluster central galaxies with colors similar to cluster red sequence galaxies. Their colors and magnitudes suggest stellar masses around $10^9 \mathrm{M_{\odot}}$. More than half of these galaxies have half-light radii smaller than 200 pc, falling into the category of massive UCDs and CEs, with properties similar to M32. 
The properties are consistent with a tidal stripping origin, but we cannot rule out the possibility that they are early-formed compact objects trapped in massive dark matter halos. The 17 CLASH clusters studied in this work on average contain 2.7 of these objects in their central 0.3 Mpc and 0.6 in their central 50 kpc. Our study demonstrates the possibility of statistically characterizing UCDs/CEs with a large set of uniform imaging survey data.
\end{abstract}
\maketitle

\section{Introduction}

Ultra-compact dwarf (UCD) galaxies were first discovered in spectroscopic surveys of the Fornax cluster \citep{1999A&AS..134...75H, 2000PASA...17..227D}. These rare objects have half-light radii between 10 and 200 pc and masses between $10^6 $ M$_\odot$ and $10^9 $ M$_\odot$. The properties position them in the once unpopulated region between globular clusters (GCs) and compact ellipticals (CEs) in the galaxy mass--size diagram. UCDs are highly concentrated objects with a spherical shape --- morphologically similar to GCs and CEs. At least some UCDs contain a central supermassive black hole (see \citealp{2014Natur.513..398S, 2015MNRAS.449.1716J} for detections, and  \citealp{2011MNRAS.414L..70F, 2015MNRAS.451.3615N} for non-detections). UCDs appear not to have high dark matter fractions  \citep{2011MNRAS.414L..70F, 2013A&A...558A..14M, 2014MNRAS.444.2993F,2015MNRAS.449.1716J}.  

UCDs span a range in masses and sizes, from those of star clusters to CE galaxies, and correspondingly are likely to have a range of formation paths.
Some UCDs appear to have a star cluster origin  \citep{2002MNRAS.330..642F,2011A&A...529A.138B, 2011MNRAS.414..739N, 2012A&A...537A...3M, 2014MNRAS.444.3670P, 2015MNRAS.446.2038R}. It has often been argued that UCDs may form by tidal stripping of initially larger galaxies   \citep{2001ApJ...557L..39B, 2001ApJ...552L.105B, 2002ApJ...568L..13G, 2008MNRAS.385.2136G, 2013MNRAS.433.1997P}. Indeed, many UCDs possess properties that are consistent with being stripped cores of dwarf or normal galaxies \citep{2010ApJ...724L..64P, 2011AJ....142..199B, 2014Natur.513..398S, 2015ApJ...812...34L, 2015ApJ...802...30Z}. 

It is also possible that UCDs may have formed as compact galaxies at very early cosmic times. In the context of CE formation, \cite{2016MNRAS.456.1030W} found that simulated compact galaxies may retain their compactness if they become satellites in a massive dark matter halo. The objects are shielded from merging and become quiescent , hence providing a window for studying early universe objects. The compact galaxies studied by \cite{2016MNRAS.456.1030W} have larger masses and sizes than UCDs. Should the trend extend to the UCD mass range, one might expect UCDs to appear as satellites captured by massive dark matter halos.

These different formation scenarios would predict differences between the emergent UCD population. Yet, our census of UCDs is still rather limited, particularly so for the most massive UCDs close to the UCD and CE boundary, like the prototypical M32 galaxy \citep{2005A&A...430L..25M, 2014MNRAS.443.1151N}. Most UCDs are identified in the nearby universe. UCD searches are limited by the UCD occurrence rate and the local universe volume. One possible approach to increasing the volume is to search for UCDs at higher redshift. Yet, identifying UCDs at higher redshift is particularly difficult due to their small sizes/masses and the difficulty of confirming them with spectroscopy. The two highest redshift GC/UCD searches respectively focus on the the Abell 1689 \citep[$z$=0.18,][]{2013ApJ...775...20A} and the Abell 2744 \citep[$z$=0.308,][]{2016ApJ...831..108L} clusters. \cite{2016ApJ...831..108L} reported $\sim$150 low-luminosity/mass ($M_r'<-14.9$ and $m_{F814W}>26$ ) UCDs in Abell 2744.

In this work, we present the discovery of UCD/CE candidates at $0.2<z<0.6$ utilizing data from the {\it C}luster {\it L}ensing {\it A}nd {\it S}upernova survey with {\it H}ubble (CLASH). Independent from \cite{2016ApJ...831..108L}, we focus on a more luminous UCD brightness range where UCD/CE occurrence rate is significantly lower. Although we did not perform spectroscopic follow-up of the candidates, their density profiles and color distributions (Section~\ref{sec:spatial_color}) provide unambiguous evidence that $\sim$45 of them are genuine high redshift objects. Analyses of their masses and sizes (Section~\ref{sec:mass}) indicate that they are among the most massive UCDs or the most compact CEs discovered to date. Throughout this paper, we assume a $\Lambda$CDM cosmology with $\Omega_{\Lambda}$ of 0.7,   $\Omega_{m}$ of 0.3 and $H_0$ of 70 km\,s$^{-1}$\,Mpc$^{-1}$. 

\section{Data}

The study is based on publicly available images from the CLASH\footnote{https://archive.stsci.edu/prepds/clash/} survey originally designed to study strong lensing clusters \citep{2012ApJS..199...25P}. For this work, we have eliminated 7 CLASH clusters
 (CLJ1226+3332, MACS1423+24, RXJ1532.9+3021, MACS1931-26, RXJ2129+0005,  MACS0416-24, MACS0647+70)
because their irregular BCG star formation filaments interfere with UCD detection. One more cluster is eliminated (MACSJ0744+39) as  we require images in the F475W, F625W, and F775W filters. This study makes use of 17 clusters in total in the redshift range of 0.19-0.6 with X-ray temperatures above 5 keV.

Low-luminosity objects around the bright central galaxies (BCGs) are difficult to identify because of light profile blending. We first subtract off the bulk of the BCG light to improve object detection at cluster centers. We use the \GALFIT software package \citep{2002AJ....124..266P, 2010AJ....139.2097P} to model the light profile of the BCGs and then subtract the BCGs from the images. This procedure is performed for the CLASH 30mas F475W, F625W, and F775W images (roughly corresponding to SDSS $g$, $r$, $i$). We model the BCGs with a single Sersic profile with a flexible Sersic index. The point-spread function (PSF) is modeled from sources identified by CLASH as stars near the cluster centers with a \citet{1969A&A.....3..455M} profile.  

We then pass BCG-subtracted F775W images through \sextractor \citep{1996A&AS..117..393B} to create a catalog of sources. To evaluate object colors, we run \sextractor in dual-image mode, using the BCG-subtracted F775W images as the detection images, and making measurements from the BCG-subtracted F475W and F625W images. 

UCDs at the distance of the CLASH clusters would be nearly unresolved by {\it Hubble}; we thus select UCD candidates for further study using \sextractor's $CLASS\_STAR$ parameter from the BCG-subtracted F775W catalogs. 
We tune the SExtractor $FWHM\_SEEING$ to be 0.15 arcsec for our purposes, slightly larger than the actual PSF (Point Spread Function) widths of the CLASH objects.  We select the ``point-like" objects with $CLASS\_STAR$ above 0.9 as possible UCDs/CEs. Subsequent tests show that this criterion picks up objects with half-light radii under 0.09 arcsec (3 pixels), assuming a Sersic index of 1. The selected objects are all referred to as ``unresolved objects" for the rest of the paper, although some of them are still above the resolution limit. 


\section{Spatial and Color Distribution}

\label{sec:spatial_color}

\begin{figure}
\includegraphics[width=0.5\textwidth]{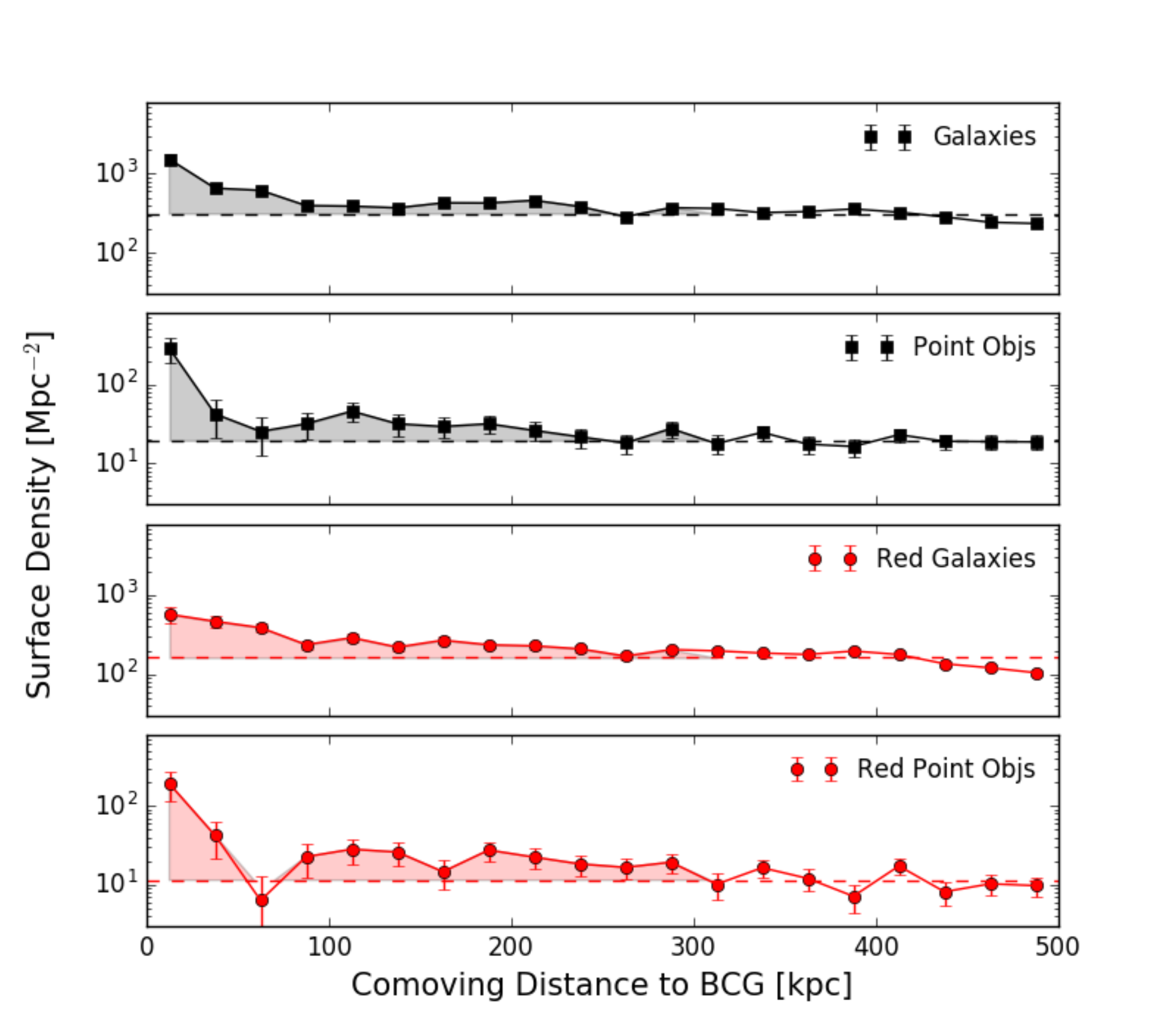}
\caption{Average surface density profiles of extended galaxies ($CLASS\_STAR <$ 0.1; top panel, black), unresolved objects ($CLASS\_STAR >$ 0.9; second panel, black), color-selected extended galaxies ($CLASS\_STAR <$ 0.1; 3rd panel, red), and color-selected unresolved objects ($CLASS\_STAR >$ 0.9; bottom panel, red). Only Poisson uncertainties are accounted for in this figure. Note that the background densities (shown as dashed lines) are not subtracted. }
\label{fig:spatial_dist}
\end{figure}

\begin{figure}
\includegraphics[width=0.5\textwidth]{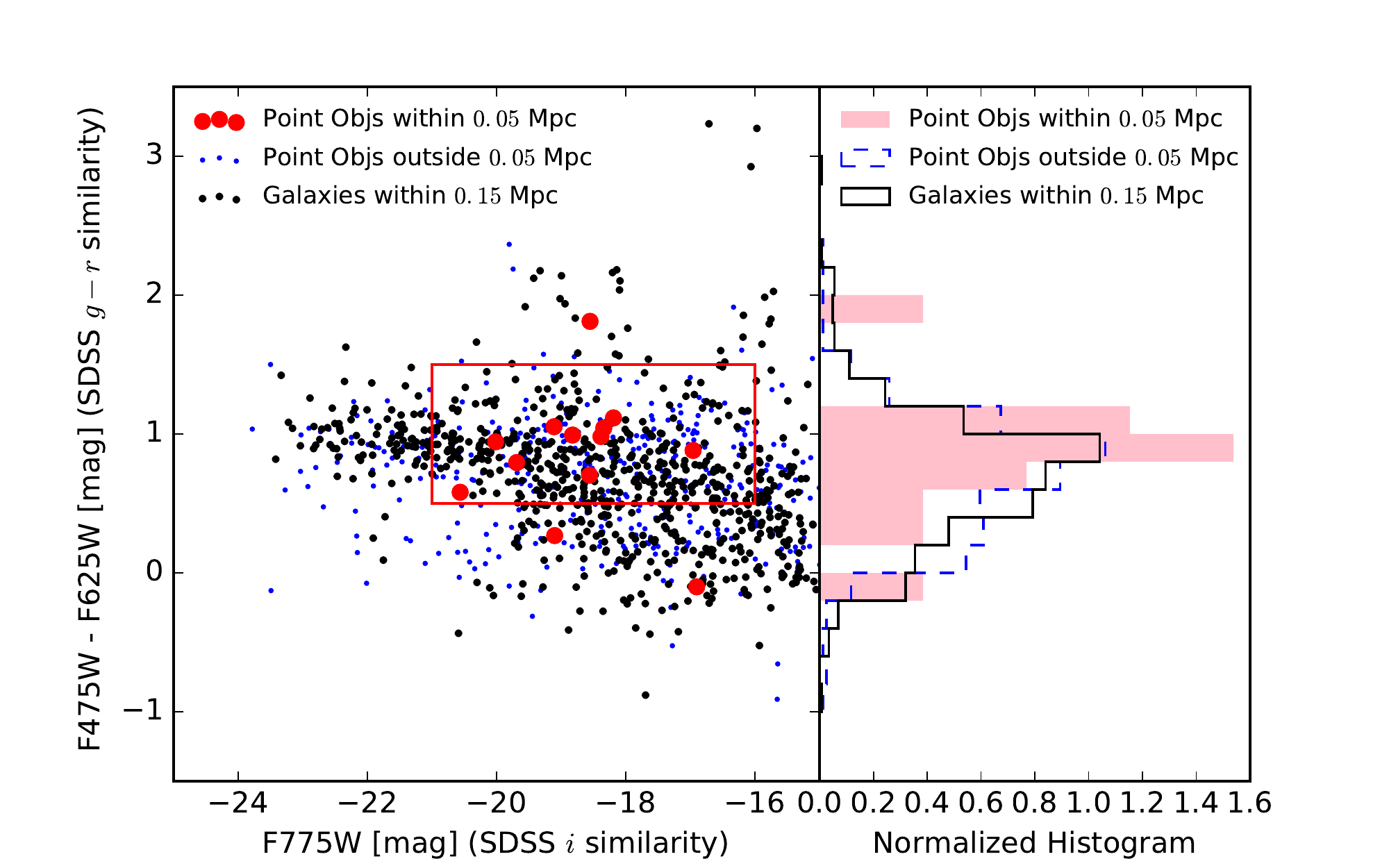}
\caption{ Rest-frame color-magnitude distribution of the unresolved objects (blue dots, $CLASS\_STAR <$ 0.9), the unresolved objects within 50 kpc of the BCGs (red solid circles), and extended galaxies ($CLASS\_STAR >$ 0.1) within 150kpc of the BCGs (black dots). The color distribution of the unresolved objects close to the BCGs is consistent with the massive end of the cluster red sequence. The box indicates the broad color selection for red galaxies used in Figure~\ref{fig:spatial_dist}.}
\label{fig:color_dist}
\end{figure}

We examine the projected density profiles of the CLASH objects in different categories. The top two panels of 
Figure~\ref{fig:spatial_dist} show the average density profiles of the extended galaxies ($CLASS\_STAR < 0.1$) and  nearly unresolved objects ($CLASS\_STAR >0.9$) in the rest-frame F775W absolute magnitude range of $-21$ to $-16$. The densities of both types increase toward the cluster centers. The density of the unresolved objects increases mildly outside the cluster core, and then sharply increases within comoving 50 kpc of the BCG centers, from 32 $ \pm$ 12 $\mathrm{Mpc}^{-2}$ in the [75 kpc, 100 kpc] radius bin to 286 $\pm$ 95 $\mathrm{Mpc}^{-2}$ within 25 kpc of the BCGs.

In order to estimate the number of UCDs/CEs in the CLASH sample, we correct for the density of foreground stars and unresolved background galaxies. Owing to the small radial coverage ($<0.5$ Mpc) of the CLASH imaging data, we choose to conservatively estimate the background density within a $0.3 - 0.5$ Mpc annulus. With such a background estimate, we infer an overall excess of 45.7 unresolved objects within a radius of 0.3 Mpc (134 versus 88.3 expected from the background). The excess of unresolved objects in the core region is even more pronounced: if there was no excess over background, we would expect only $(0.61\pm0.05)$ objects within 25 kpc of all the BCGs, or $(2.45 \pm 0.20)$ within 50 kpc. The observed numbers are 9 and 13, respectively. Assuming a Poisson distribution, the probability of observing 9 or more unresolved objects within 25 kpc is only  $2\times10^{-8}$, and the probability of observing 13 or more unresolved objects within 50 kpc is only $2 \times 10^{-6}$.  For comparison, the probability of observing a 5 $\sigma$ outlier event in a Gaussian model is $6\times10^{-7}$, or $7\times10^{-6}$ for a 4.5 $\sigma$ event. The significant overdensity of the unresolved objects is strong evidence that a portion of them are associated with the clusters.
 
The projected density of extended galaxies increases toward the center already at 500kpc, steepening somewhat toward the cluster centers.  The density increase of the unresolved objects may be more dramatic than that of the extended galaxies. The density of unresolved objects within 25kpc is $9.0\pm 4.5$ times the density measured between 100 and 125 kpc, a factor of more than two higher than the increase of a factor of $4\pm 1$ seen for extended galaxies (but with large uncertainties that admit an acceptable fit with no difference between the samples).

Further insight can be gleaned by examining the observed colors of the objects within 50kpc of the cluster centers. Unresolved objects within 50kpc of the cluster centers have a similar color distribution to that of the extended galaxies (which are predominantly at the cluster redshift). This suggests that the vast majority of the unresolved objects within 50kpc are indeed galaxies at the redshift of the clusters. In contrast, unresolved objects drawn from across the field have a wider range of observed colors, consistent with our argument that while some of them are UCDs/CEs at the cluster redshift, many of them are foreground stars or background unresolved sources.

We then examine the distribution of the selected objects in the rest-frame color-magnitude diagram. For each cluster field, we compute $K$-corrections following \cite{2007AJ....133..734B}, assuming that all objects are located at the cluster redshifts. 
In Fig.~\ref{fig:color_dist}, we show the rest-frame colors and absolute magnitudes of the extended galaxies close to the cluster centers (black data points), the unresolved objects within 50kpc of the cluster centers (red filled circles), and all the rest of the unresolved objects (blue data points). The color-magnitude distribution of the extended galaxies demonstrates a clear red sequence. The colors of the unresolved objects inside the cores are consistent with the bright end of the red sequence, providing further evidence that these objects are associated with the cluster galaxy population. The unresolved objects outside the cluster cores show a peak at red colors, but have a wider color distribution, consistent with the idea that many of them are foreground stars or background unresolved sources. Taken together, the balance of evidence strongly suggests that many of the unresolved sources detected in this study are compact galaxies at the cluster redshift; in particular, 80\% of such sources within 50 kpc are expected to be cluster galaxies. 

In the bottom two panels of Fig.~\ref{fig:spatial_dist}, we further restrict the selection of galaxies and unresolved sources by imposing a rest-frame color cut of $0.5<F475W - F625W < 1.5$. The surface density increase of the unresolved objects is still most noticeable within 50 kpc. The majority of the unresolved objects within 50 kpc of the cluster centers are red objects.

\section{Mass and Size estimates}
\label{sec:mass}

\begin{table*}
\begin{center}
\caption{UCD/CE Candidate List \label{tbl:table}}
\begin{tabular}{ccccccccc}
\tableline\tableline 
\\
   & Designation & Host Cluster & Redshift & R.A.  & Decl. & \GALFIT $Re$ (pc) & $CLASS\_STAR$ $Re$  & F775W Magnitude \\
& & &(Cluster)  & (J2000) &(J2000)  &[pc]  & Limit [pc]  &  0$\farcs$167, diameter \\
& & & & & & & & extinction corrected \\
\\
 & CLASHJ024803.4-033145.8 & a383 & 0.187 &        42.014095 &       $-$3.5293905 & 240$\pm$7$^*$ &  $<$153 & 21.28\\
 & CLASHJ032941.9-021146.3 & macs0329 & 0.450 &        52.424718 &       $-$2.1961972 & 164$\pm$18 &  $<$344 & 24.83\\
 & CLASHJ042935.9-025310.7 & macs0429 & 0.399 &        67.399641 &       $-$2.8863028 & 47$\pm$23 &  $<$397 & 23.03\\
 & CLASHJ042936.0-025308.5 & macs0429 & 0.399 &        67.399941 &       $-$2.8856861 & 271$\pm$18 &  $<$397 & 24.67\\
 & CLASHJ115717.2+333642.8 & a1423 & 0.213 &        179.32166 &        33.611894 & N.A. &  $<$165 & 24.14\\
 & CLASHJ120612.5-084802.1 & macs1206 & 0.440 &        181.55214 &       $-$8.8005804 & 103$\pm$13 &  $<$387 & 23.77\\
 & CLASHJ172016.6+353624.2 & macs1720 & 0.391 &        260.06923 &        35.606736 & 213$\pm$10 &  $<$384 & 23.95\\
 & CLASHJ172016.7+353629.5 & macs1720 & 0.391 &        260.06941 &        35.608194 & 102$\pm$17 &  $<$384 & 24.61\\
 & CLASHJ172016.6+353632.6 & macs1720 & 0.391 &        260.06918 &        35.609061 & N.A. &  $<$384 & 25.94\\
 & CLASHJ172226.4+320754.4 & a2261 & 0.224 &        260.60991 &        32.131771 & N.A. &  $<$173 & 20.47\\
 & CLASHJ172227.1+320755.8 & a2261 & 0.224 &        260.61306 &        32.132179 & 54$\pm$27 &  $<$173 & 23.70\\
 & CLASHJ172227.2+320757.7 & a2261 & 0.224 &        260.61346 &        32.132688 & 125$\pm$12 &  $<$173 & 22.64\\
 & CLASHJ172227.2+320757.6 & a2261 & 0.224 &        260.61335 &        32.132654 & 161$\pm$59 &  $<$173 & 23.11\\
\tableline
\end{tabular}
\end{center}
\vspace{-1em}
 $^*$ This object appears to be a bright star. The \GALFIT $Re$ is higher than the $CLASS\_STAR$ constraint, possibly because of saturation.
\end{table*}

\begin{figure*}
\includegraphics[width=0.9\textwidth]{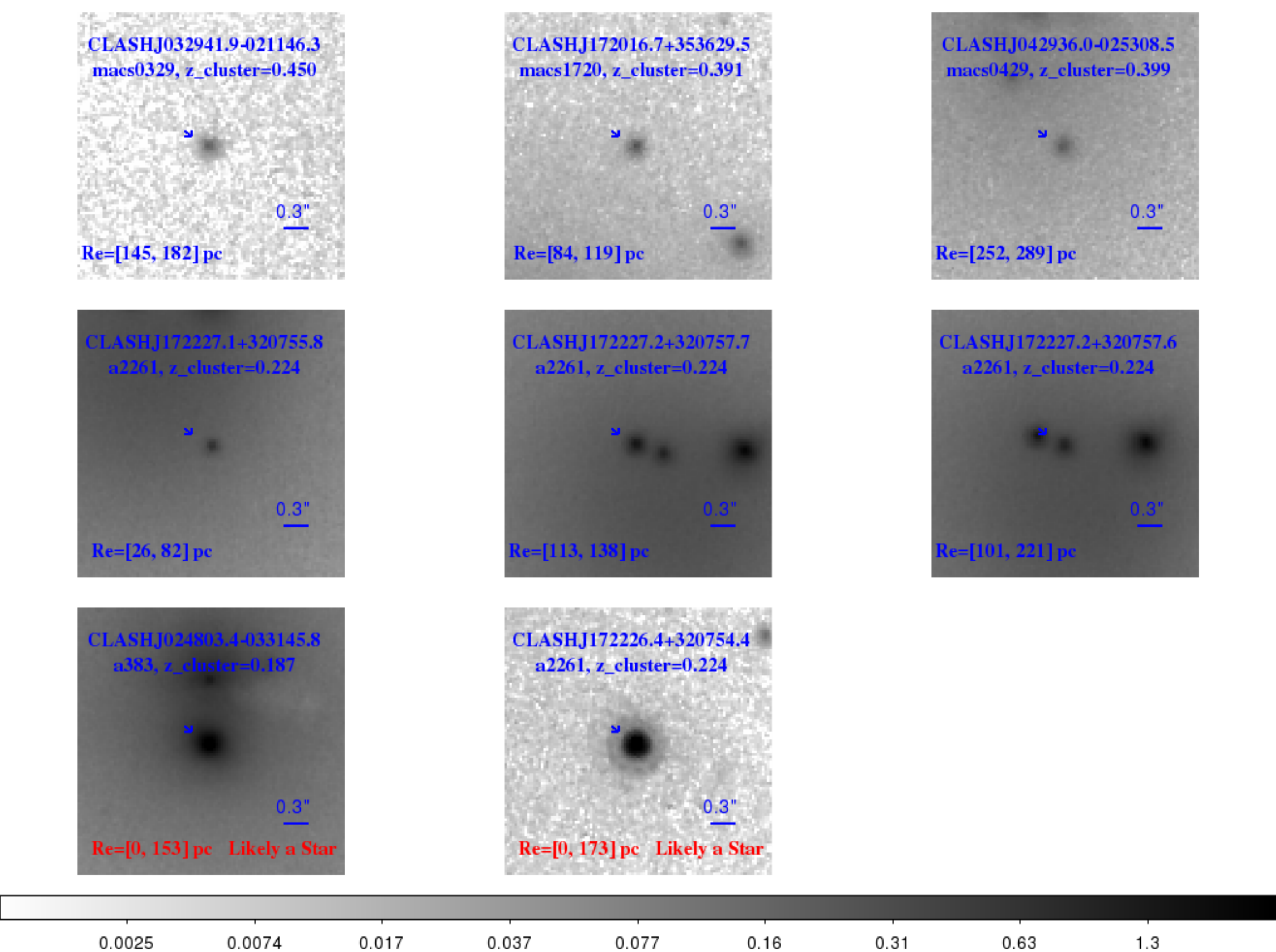}
\caption{Images, designation, and host information of eight UCD/CE candidates in the CLASH clusters. In the bottom panels, we also show two candidates that are likely foreground stars. Comparison with the density of foreground/background unresolved sources suggests that $\sim80\%$ of the sample within 50kpc are expected to be real UCDs/CEs. }
\label{fig:list}
\end{figure*}

\begin{figure*}
\includegraphics[width=1.0\textwidth]{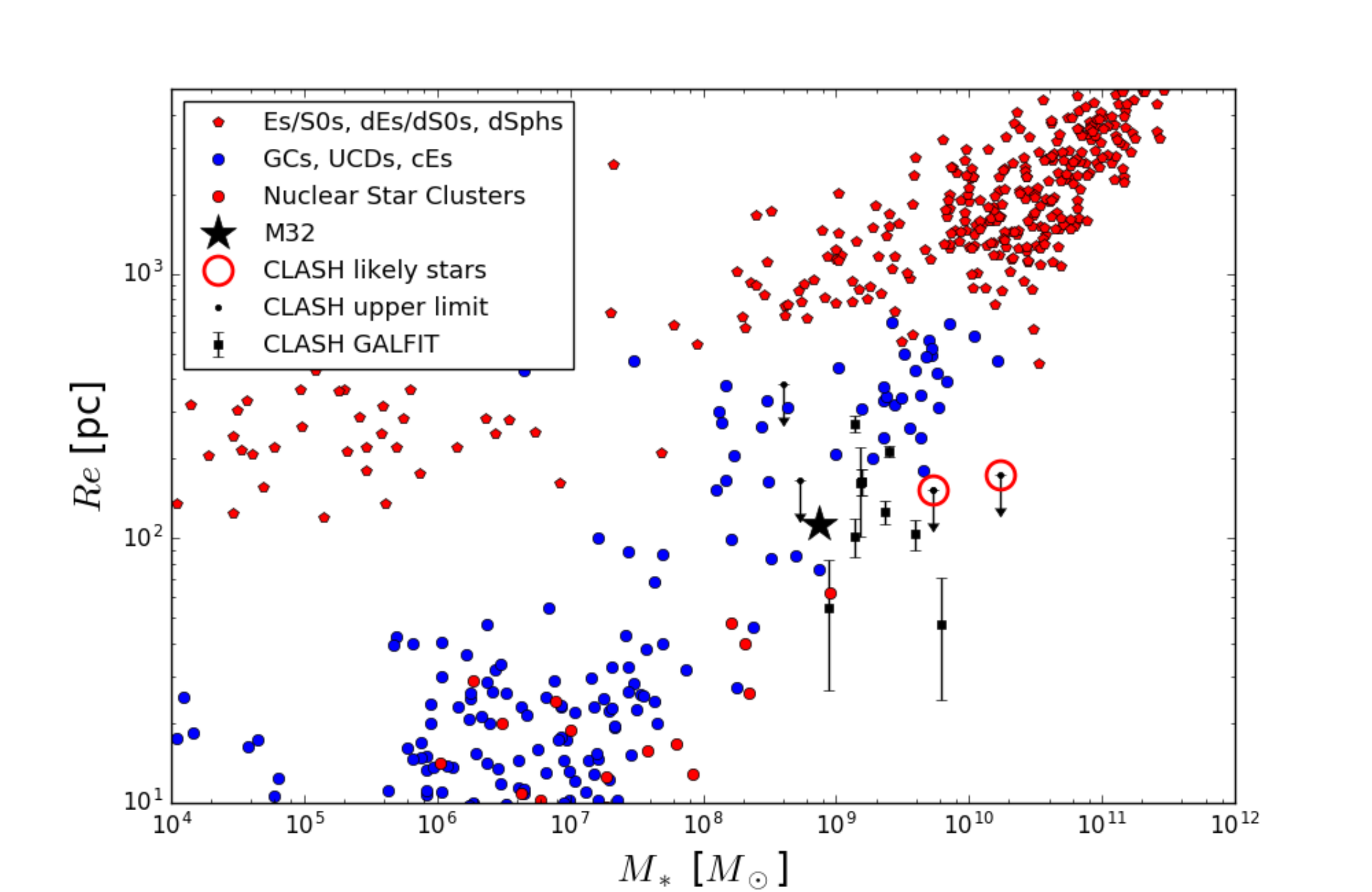}
\caption{The mass and size distribution of the UCD/CE candidates reported in this paper. For the UCD/CE half-light radii, we plot either the \GALFIT values or the $CLASS\_STAR$ upper limits, whichever is lower. The two likely stars reported in Figure~\ref{fig:list} are indicated by red circles. We also overplot the list of nuclei, GCs, UCDs, CEs and dwarf galaxies in \cite{2014MNRAS.443.1151N}. The UCD/CE satellite of the Andromeda galaxy, M32, is marked with a filled star. The UCD/CE population reported in this paper appears to be analogous to M32.}
\label{fig:mass_size}
\end{figure*}

In Section~\ref{sec:spatial_color}, we identify an interesting population of unresolved objects that are in close proximity to the cluster centers and have colors consistent with cluster galaxies. The half-light radii of these objects are below 0.09 arcsec (3 pixels) as they are unresolved, which places them in the size regime of GCs, dwarf galaxies, and compact galaxies. The galaxy candidates have $F775W$-band magnitudes of 22-25 mag in the redshift range of $\sim$0.2-$\sim$0.6.

We estimate their stellar masses from 0$\farcs$167 aperture magnitudes using the \citet{2003MNRAS.344.1000B} stellar population synthesis model with a \citet{1955ApJ...121..161S} stellar initial mass function, following  model \citet{2012PASP..124..606M}. We adopt a formation redshift of 3.0, with an exponentially decaying star formation history with $\tau = 0.1 \mathrm{Gyrs}$ and $Z=0.008$. The resultant  stellar masses are above $10^8\mathrm{M_{\odot}}$, which indicates extremely high stellar density. Their mass and size distributions are consistent with being UCD or CE galaxies.

In this section, we further examine the properties of the 13 unresolved objects within 50 kpc of the BCG centers; these span a slightly lower redshift range of $0.2<z<0.5$. We choose to focus on these objects because of the low rate of background contamination in the core (19\% contamination within 50 kpc instead of 66\% within 0.3 Mpc). Images, designation, and host information of the selected objects are listed in Table~\ref{tbl:table}. None of these 13 objects appear to be previously reported as UCD/CE candidates in literature.{\footnote{Search performed on https://ned.ipac.caltech.edu.}} It is likely that some of these objects are foreground stars, but from the excess of their surface density (Figure~\ref{fig:spatial_dist}), we infer that $ \sim 80\%$ of them are real UCDs/CEs. We show postage images of eight candidates --- six likely UCDs/CEs and two that we consider more likely to be foreground stars --- in Figure~\ref{fig:list}. 

In Figure ~\ref{fig:mass_size}, we show the 13 candidate UCDs/CEs in the mass--size diagram. For 10 of the 13 candidates, we estimate half-light radii ($Re$) with \GALFIT, using a single Sersic profile. Owing to the small size of the candidates, we choose to fix the Sersic index to $n$ = 1 given that many local UCDs/CEs have $n\sim 1-2$. \citep{2008AJ....136.2295B, 2008AJ....136..461E}. We note that the adoption of $n$ = 4 results in unsatisfactory fits.  We are unable to estimate $Re$ for three objects. Their $Re$s are likely to be below 0.015 arcsec(0.5 pixel) and \GALFIT does not converge. 

For all 13 objects, we estimate upper limits to their half-light radii using their $CLASS\_STAR$ in concert with simulations. We inject faint $24<F775W<25$ PSF-convolved $n=1$ profiles with a range of half-light radii below 0.3 arcsec (10 pixels) into the F775W images. We run \sextractor on the simulated images and match the detected objects to their true positions, magnitudes and half-light radii. 
The $CLASS\_STAR$ measurements yield higher $CLASS\_STAR$ values with decreasing radii.  For example, in the cluster field of MACS1720, $CLASS\_STAR$ increases from 0 to 1 when the simulation $Re$s drop from 4 to 2 pixels. For each cluster field, we select simulated objects with $CLASS\_STAR$ above 0.9, and on this basis estimate a 84.13 percentile of the true $Re$ values (the percentile that correspond to a $1\sigma$ outlier event in a Gaussian distribution), to be a 1$\sigma$ upper limit to $Re$ for the UCD/CE candidates.

Figure ~\ref{fig:mass_size} shows the smaller of either the \GALFIT or the $CLASS\_STAR$-derived $Re$ upper limits for the UCD/CE candidates. The half-light radii of all candidate UCDs/CEs are below 400 pc. Eight candidates have $Re$ below 200 pc. In this figure, we also plot the the masses and sizes of the UCDs, CEs, GCs, nuclei and dwarf galaxies compiled in \cite{2014MNRAS.443.1151N}. 
In this paper, the reported candidates further populate the sparse region between GCs and CEs. Comparing to the identified UCDs in the literature, the sample reported in this work occupies the most massive UCD and the most compact CE region in the mass and size diagram.

In Figure~\ref{fig:mass_size}, it becomes apparent that the UCD/CE candidates are analogous to M32 -- a satellite of the Andromeda galaxy and one of the most massive/compact UCDs/CEs. At least four objects reported in this paper appear strikingly similar to M32 in terms of their masses and  sizes. Some candidates are more compact than M32, even after  considering a 0.3 dex stellar mass uncertainty \citep[e.g., using a different IMF model will result in a 0.3 dex stellar mass difference][]{2003PASP..115..763C}.

\section{Discussion and Summary}

In this paper, we discovered a sample of unresolved objects in 17 $0.2<z<0.6$ CLASH clusters. They are strongly clustered around the BCGs, and have inferred masses and colors consistent with being UCD/CE galaxies. Although the sample is not spectroscopically confirmed, its density profile and color distribution present unambiguous evidence that some of the objects are truly high redshift UCDs/CEs. 

The discovered candidates appear in the massive end of the UCD mass range, yet are strikingly compact, with half of the sample having sizes below 200 pc. The masses and sizes indicate that this sample contains some of the densest UCDs/CEs discovered to date, with a number of them denser than the M32 UCD/CE satellite of the Andromeda galaxy.  

Estimating the background density 0.3--0.5 Mpc from the cluster centers, we infer an overall excess of 45.7 UCD and CE-like objects inside 0.3 Mpc for all the 17 CLASH clusters, or an excess of 10.5 objects within 50 kpc. One CLASH cluster on average contains $\sim 0.6$ massive UCD/CE within 50 kpc of the BCG, or $\sim 2.7$ massive UCDs/CEs within 0.3 Mpc. Interestingly, these objects appear to be more common in CLASH clusters than the compact objects in the local universe ($\sim 10^{-6}$ Mpc$^{-3}$, N. Trujillo in prep., private communication). Nevertheless, the occurrence rate of the massive UCDs or CEs is low. Furthermore, recognizing them is not straightforward due to easy confusion with foreground stars. These together explain the rarity of massive UCD or CE discoveries in nearby clusters.

The colors, masses, and spatial distributions of these candidates have a bearing on possible formation routes for this sample. Given the mass range of the reported sample, these UCD/CE candidates are unlikely to be massive GCs \citep{2011MNRAS.414..739N}. Because many of the candidates are concentrated toward the cluster core, the reported sample is consistent with being stripped cores of cluster galaxies. Early-type galaxies tend to have redder centers, due to older stellar ages and higher stellar metallicities \citep{2015ApJ...807...11G}. Most of the reported UCD/CE candidates in cluster cores reside on the redder side of the red sequence. Should the stripping origin be verified, these massive UCDs or CEs demonstrate that stripping toward forming compact objects \citep[e.g., see][]{2015Sci...348..418C} is ubiquitous across a wide mass range. Yet, we cannot rule out the possibility that the UCDs/CEs are early-formed compact objects trapped inside galaxy clusters \citep{2016MNRAS.456.1030W} --- such an origin would also be consistent with red colors and a centrally concentrated distribution. More precise analyses of the UCD/CE spatial and color distributions with a bigger sample, in concert with guidance from simulations, may help to distinguish between possible formation channels. For example, with a stripping origin, we should observe a higher concentration of UCDs/CEs compared to cluster galaxies. In this paper, the UCD/CE sample appears to be more concentrated than the extended galaxies by roughly a factor of two, corresponding to a $\sim$ 1 $\sigma$ difference. A $\sim 9\times$ larger sample would provide better number statistics, and if this concentration difference remains it would be detected at $\sim 3\sigma$. 

It may be possible to apply the statistical approach used in this work to Sloan Digital Sky Survey or Dark Energy Survey data. Due to blending between compact objects and luminous galaxies, we may need to apply advanced deblending techniques \citep{2012MNRAS.422..449B, 2013ApJS..206...10G, 2015PASP..127.1183Z} to these ground-based survey data. Analyses based on the wide-field surveys should yield much more precise statistical measurements of UCD or CE properties, therefore strongly constraining the mechanisms of the formation of compact objects.

\acknowledgements
Y. Zhang acknowledges supports by the University of Michigan Rackham Pre-Doctoral fellowship and the Fermilab Schramm Post-Doctoral fellowship. We thank Prof. Timothy McKay, Dr. Nacho Trujillo, Prof. Michael Fellhauer, and Prof. Alister Graham for helpful discussions and insightful comments.


\begin{thebibliography}{45}
\expandafter\ifx\csname natexlab\endcsname\relax\def\natexlab#1{#1}\fi

\bibitem[{{Alamo-Mart{\'{\i}}nez} {et~al.}(2013){Alamo-Mart{\'{\i}}nez},
  {Blakeslee}, {Jee}, {C{\^o}t{\'e}}, {Ferrarese},
  {Gonz{\'a}lez-L{\'o}pezlira}, {Jord{\'a}n}, {Meurer}, {Peng}, \&
  {West}}]{2013ApJ...775...20A}
{Alamo-Mart{\'{\i}}nez}, K.~A., {Blakeslee}, J.~P., {Jee}, M.~J., {et~al.}
  2013, \apj, 775, 20

\bibitem[{{Barden} {et~al.}(2012){Barden}, {H{\"a}u{\ss}ler}, {Peng},
  {McIntosh}, \& {Guo}}]{2012MNRAS.422..449B}
{Barden}, M., {H{\"a}u{\ss}ler}, B., {Peng}, C.~Y., {McIntosh}, D.~H., \&
  {Guo}, Y. 2012, \mnras, 422, 449

\bibitem[{{Bekki} {et~al.}(2001{\natexlab{a}}){Bekki}, {Couch}, \&
  {Drinkwater}}]{2001ApJ...552L.105B}
{Bekki}, K., {Couch}, W.~J., \& {Drinkwater}, M.~J. 2001{\natexlab{a}}, \apjl,
  552, L105

\bibitem[{{Bekki} {et~al.}(2001{\natexlab{b}}){Bekki}, {Couch}, {Drinkwater},
  \& {Gregg}}]{2001ApJ...557L..39B}
{Bekki}, K., {Couch}, W.~J., {Drinkwater}, M.~J., \& {Gregg}, M.~D.
  2001{\natexlab{b}}, \apjl, 557, L39

\bibitem[{{Bertin} \& {Arnouts}(1996)}]{1996A&AS..117..393B}
{Bertin}, E., \& {Arnouts}, S. 1996, \aaps, 117, 393

\bibitem[{{Blakeslee} \& {Barber DeGraaff}(2008)}]{2008AJ....136.2295B}
{Blakeslee}, J.~P., \& {Barber DeGraaff}, R. 2008, \aj, 136, 2295

\bibitem[{{Blanton} \& {Roweis}(2007)}]{2007AJ....133..734B}
{Blanton}, M.~R., \& {Roweis}, S. 2007, \aj, 133, 734

\bibitem[{{Brodie} {et~al.}(2011){Brodie}, {Romanowsky}, {Strader}, \&
  {Forbes}}]{2011AJ....142..199B}
{Brodie}, J.~P., {Romanowsky}, A.~J., {Strader}, J., \& {Forbes}, D.~A. 2011,
  \aj, 142, 199

\bibitem[{{Br{\"u}ns} {et~al.}(2011){Br{\"u}ns}, {Kroupa}, {Fellhauer}, {Metz},
  \& {Assmann}}]{2011A&A...529A.138B}
{Br{\"u}ns}, R.~C., {Kroupa}, P., {Fellhauer}, M., {Metz}, M., \& {Assmann}, P.
  2011, \aap, 529, A138

\bibitem[{{Bruzual} \& {Charlot}(2003)}]{2003MNRAS.344.1000B}
{Bruzual}, G., \& {Charlot}, S. 2003, \mnras, 344, 1000

\bibitem[{{Chabrier}(2003)}]{2003PASP..115..763C}
{Chabrier}, G. 2003, \pasp, 115, 763

\bibitem[{{Chilingarian} \& {Zolotukhin}(2015)}]{2015Sci...348..418C}
{Chilingarian}, I., \& {Zolotukhin}, I. 2015, Science, 348, 418

\bibitem[{{Drinkwater} {et~al.}(2000){Drinkwater}, {Jones}, {Gregg}, \&
  {Phillipps}}]{2000PASA...17..227D}
{Drinkwater}, M.~J., {Jones}, J.~B., {Gregg}, M.~D., \& {Phillipps}, S. 2000,
  pasa, 17, 227

\bibitem[{{Evstigneeva} {et~al.}(2008){Evstigneeva}, {Drinkwater}, {Peng},
  {Hilker}, {De Propris}, {Jones}, {Phillipps}, {Gregg}, \&
  {Karick}}]{2008AJ....136..461E}
{Evstigneeva}, E.~A., {Drinkwater}, M.~J., {Peng}, C.~Y., {et~al.} 2008, \aj,
  136, 461

\bibitem[{{Fellhauer} \& {Kroupa}(2002)}]{2002MNRAS.330..642F}
{Fellhauer}, M., \& {Kroupa}, P. 2002, \mnras, 330, 642

\bibitem[{{Forbes} {et~al.}(2014){Forbes}, {Norris}, {Strader}, {Romanowsky},
  {Pota}, {Kannappan}, {Brodie}, \& {Huxor}}]{2014MNRAS.444.2993F}
{Forbes}, D.~A., {Norris}, M.~A., {Strader}, J., {et~al.} 2014, \mnras, 444,
  2993

\bibitem[{{Frank} {et~al.}(2011){Frank}, {Hilker}, {Mieske}, {Baumgardt},
  {Grebel}, \& {Infante}}]{2011MNRAS.414L..70F}
{Frank}, M.~J., {Hilker}, M., {Mieske}, S., {et~al.} 2011, \mnras, 414, L70

\bibitem[{{Galametz} {et~al.}(2013){Galametz}, {Grazian}, {Fontana},
  {Ferguson}, {Ashby}, {Barro}, {Castellano}, {Dahlen}, {Donley}, {Faber},
  {Grogin}, {Guo}, {Huang}, {Kocevski}, {Koekemoer}, {Lee}, {McGrath}, {Peth},
  {Willner}, {Almaini}, {Cooper}, {Cooray}, {Conselice}, {Dickinson}, {Dunlop},
  {Fazio}, {Foucaud}, {Gardner}, {Giavalisco}, {Hathi}, {Hartley}, {Koo},
  {Lai}, {de Mello}, {McLure}, {Lucas}, {Paris}, {Pentericci}, {Santini},
  {Simpson}, {Sommariva}, {Targett}, {Weiner}, {Wuyts}, \& {the CANDELS
  Team}}]{2013ApJS..206...10G}
{Galametz}, A., {Grazian}, A., {Fontana}, A., {et~al.} 2013, \apjs, 206, 10

\bibitem[{{Goerdt} {et~al.}(2008){Goerdt}, {Moore}, {Kazantzidis}, {Kaufmann},
  {Macci{\`o}}, \& {Stadel}}]{2008MNRAS.385.2136G}
{Goerdt}, T., {Moore}, B., {Kazantzidis}, S., {et~al.} 2008, \mnras, 385, 2136

\bibitem[{{Graham}(2002)}]{2002ApJ...568L..13G}
{Graham}, A.~W. 2002, \apjl, 568, L13

\bibitem[{{Greene} {et~al.}(2015){Greene}, {Janish}, {Ma}, {McConnell},
  {Blakeslee}, {Thomas}, \& {Murphy}}]{2015ApJ...807...11G}
{Greene}, J.~E., {Janish}, R., {Ma}, C.-P., {et~al.} 2015, \apj, 807, 11

\bibitem[{{Hilker} {et~al.}(1999){Hilker}, {Infante}, {Vieira},
  {Kissler-Patig}, \& {Richtler}}]{1999A&AS..134...75H}
{Hilker}, M., {Infante}, L., {Vieira}, G., {Kissler-Patig}, M., \& {Richtler},
  T. 1999, \aaps, 134, 75

\bibitem[{{Janz} {et~al.}(2015){Janz}, {Forbes}, {Norris}, {Strader}, {Penny},
  {Fagioli}, \& {Romanowsky}}]{2015MNRAS.449.1716J}
{Janz}, J., {Forbes}, D.~A., {Norris}, M.~A., {et~al.} 2015, \mnras, 449, 1716

\bibitem[{{Lee} \& {Jang}(2016)}]{2016ApJ...831..108L}
{Lee}, M.~G., \& {Jang}, I.~S. 2016, \apj, 831, 108

\bibitem[{{Liu} {et~al.}(2015){Liu}, {Peng}, {C{\^o}t{\'e}}, {Ferrarese},
  {Jord{\'a}n}, {Mihos}, {Zhang}, {Mu{\~n}oz}, {Puzia}, {Lan{\c c}on}, {Gwyn},
  {Cuillandre}, {Blakeslee}, {Boselli}, {Durrell}, {Duc}, {Guhathakurta},
  {MacArthur}, {Mei}, {S{\'a}nchez-Janssen}, \& {Xu}}]{2015ApJ...812...34L}
{Liu}, C., {Peng}, E.~W., {C{\^o}t{\'e}}, P., {et~al.} 2015, \apj, 812, 34

\bibitem[{{Mancone} \& {Gonzalez}(2012)}]{2012PASP..124..606M}
{Mancone}, C.~L., \& {Gonzalez}, A.~H. 2012, \pasp, 124, 606

\bibitem[{{Mieske} {et~al.}(2013){Mieske}, {Frank}, {Baumgardt},
  {L{\"u}tzgendorf}, {Neumayer}, \& {Hilker}}]{2013A&A...558A..14M}
{Mieske}, S., {Frank}, M.~J., {Baumgardt}, H., {et~al.} 2013, \aap, 558, A14

\bibitem[{{Mieske} {et~al.}(2012){Mieske}, {Hilker}, \&
  {Misgeld}}]{2012A&A...537A...3M}
{Mieske}, S., {Hilker}, M., \& {Misgeld}, I. 2012, \aap, 537, A3

\bibitem[{{Mieske} {et~al.}(2005){Mieske}, {Infante}, {Hilker}, {Hertling},
  {Blakeslee}, {Ben{\'{\i}}tez}, {Ford}, \& {Zekser}}]{2005A&A...430L..25M}
{Mieske}, S., {Infante}, L., {Hilker}, M., {et~al.} 2005, \aap, 430, L25

\bibitem[{{Moffat}(1969)}]{1969A&A.....3..455M}
{Moffat}, A.~F.~J. 1969, \aap, 3, 455

\bibitem[{{Norris} {et~al.}(2015){Norris}, {Escudero}, {Faifer}, {Kannappan},
  {Forte}, \& {van den Bosch}}]{2015MNRAS.451.3615N}
{Norris}, M.~A., {Escudero}, C.~G., {Faifer}, F.~R., {et~al.} 2015, \mnras,
  451, 3615

\bibitem[{{Norris} \& {Kannappan}(2011)}]{2011MNRAS.414..739N}
{Norris}, M.~A., \& {Kannappan}, S.~J. 2011, \mnras, 414, 739

\bibitem[{{Norris} {et~al.}(2014){Norris}, {Kannappan}, {Forbes}, {Romanowsky},
  {Brodie}, {Faifer}, {Huxor}, {Maraston}, {Moffett}, {Penny}, {Pota},
  {Smith-Castelli}, {Strader}, {Bradley}, {Eckert}, {Fohring}, {McBride},
  {Stark}, \& {Vaduvescu}}]{2014MNRAS.443.1151N}
{Norris}, M.~A., {Kannappan}, S.~J., {Forbes}, D.~A., {et~al.} 2014, \mnras,
  443, 1151

\bibitem[{{Paudel} {et~al.}(2010){Paudel}, {Lisker}, \&
  {Janz}}]{2010ApJ...724L..64P}
{Paudel}, S., {Lisker}, T., \& {Janz}, J. 2010, \apjl, 724, L64

\bibitem[{{Peng} {et~al.}(2002){Peng}, {Ho}, {Impey}, \&
  {Rix}}]{2002AJ....124..266P}
{Peng}, C.~Y., {Ho}, L.~C., {Impey}, C.~D., \& {Rix}, H.-W. 2002, \aj, 124, 266

\bibitem[{{Peng} {et~al.}(2010){Peng}, {Ho}, {Impey}, \&
  {Rix}}]{2010AJ....139.2097P}
---. 2010, \aj, 139, 2097

\bibitem[{{Pfeffer} \& {Baumgardt}(2013)}]{2013MNRAS.433.1997P}
{Pfeffer}, J., \& {Baumgardt}, H. 2013, \mnras, 433, 1997

\bibitem[{{Pfeffer} {et~al.}(2014){Pfeffer}, {Griffen}, {Baumgardt}, \&
  {Hilker}}]{2014MNRAS.444.3670P}
{Pfeffer}, J., {Griffen}, B.~F., {Baumgardt}, H., \& {Hilker}, M. 2014, \mnras,
  444, 3670

\bibitem[{{Postman} {et~al.}(2012){Postman}, {Coe}, {Ben{\'{\i}}tez},
  {Bradley}, {Broadhurst}, {Donahue}, {Ford}, {Graur}, {Graves}, {Jouvel},
  {Koekemoer}, {Lemze}, {Medezinski}, {Molino}, {Moustakas}, {Ogaz}, {Riess},
  {Rodney}, {Rosati}, {Umetsu}, {Zheng}, {Zitrin}, {Bartelmann}, {Bouwens},
  {Czakon}, {Golwala}, {Host}, {Infante}, {Jha}, {Jimenez-Teja}, {Kelson},
  {Lahav}, {Lazkoz}, {Maoz}, {McCully}, {Melchior}, {Meneghetti}, {Merten},
  {Moustakas}, {Nonino}, {Patel}, {Reg{\"o}s}, {Sayers}, {Seitz}, \& {Van der
  Wel}}]{2012ApJS..199...25P}
{Postman}, M., {Coe}, D., {Ben{\'{\i}}tez}, N., {et~al.} 2012, \apjs, 199, 25

\bibitem[{{Renaud} {et~al.}(2015){Renaud}, {Bournaud}, \&
  {Duc}}]{2015MNRAS.446.2038R}
{Renaud}, F., {Bournaud}, F., \& {Duc}, P.-A. 2015, \mnras, 446, 2038

\bibitem[{{Salpeter}(1955)}]{1955ApJ...121..161S}
{Salpeter}, E.~E. 1955, \apj, 121, 161

\bibitem[{{Seth} {et~al.}(2014){Seth}, {van den Bosch}, {Mieske}, {Baumgardt},
  {Brok}, {Strader}, {Neumayer}, {Chilingarian}, {Hilker}, {McDermid},
  {Spitler}, {Brodie}, {Frank}, \& {Walsh}}]{2014Natur.513..398S}
{Seth}, A.~C., {van den Bosch}, R., {Mieske}, S., {et~al.} 2014, \nat, 513, 398

\bibitem[{{Wellons} {et~al.}(2016){Wellons}, {Torrey}, {Ma}, {Rodriguez-Gomez},
  {Pillepich}, {Nelson}, {Genel}, {Vogelsberger}, \&
  {Hernquist}}]{2016MNRAS.456.1030W}
{Wellons}, S., {Torrey}, P., {Ma}, C.-P., {et~al.} 2016, \mnras, 456, 1030

\bibitem[{{Zhang} {et~al.}(2015{\natexlab{a}}){Zhang}, {Peng}, {C{\^o}t{\'e}},
  {Liu}, {Ferrarese}, {Cuillandre}, {Caldwell}, {Gwyn}, {Jord{\'a}n}, {Lan{\c
  c}on}, {Li}, {Mu{\~n}oz}, {Puzia}, {Bekki}, {Blakeslee}, {Boselli},
  {Drinkwater}, {Duc}, {Durrell}, {Emsellem}, {Firth}, \&
  {S{\'a}nchez-Janssen}}]{2015ApJ...802...30Z}
{Zhang}, H.-X., {Peng}, E.~W., {C{\^o}t{\'e}}, P., {et~al.} 2015{\natexlab{a}},
  \apj, 802, 30

\bibitem[{{Zhang} {et~al.}(2015{\natexlab{b}}){Zhang}, {McKay}, {Bertin},
  {Jeltema}, {Miller}, {Rykoff}, \& {Song}}]{2015PASP..127.1183Z}
{Zhang}, Y., {McKay}, T.~A., {Bertin}, E., {et~al.} 2015{\natexlab{b}}, \pasp,
  127, 1183

\end{thebibliography}

\end{document}